\documentstyle[sprocl]{article}

\begin{document} 

\title{GLUON DEPLETION AND $J/\psi$
SUPPRESSION \\IN $pA$ COLLISIONS}
\author{Rudolph C. Hwa }
\vskip.5cm
 
\address{{Institute of Theoretical Science and Department of
Physics\\ University of Oregon, Eugene, OR 97403-5203,
USA}}

\maketitle\abstracts{The gluon distribution of a nucleon
propagating through a nucleus can change depending on how
far the penetration depth is. It is a nonperturbative process
that we describe by an evolution equation. The kernel of the
integral equation is to be determined by a phenomenological
study of $J/\psi$ suppression in pA collision, treated as a
perturbative process. The data of E866 on $\alpha(x_F)$
shows significant dependence on $x_F$ at large $x_F$. It
presents a feature that has not been explained by any
dynamical model. We show that gluon depletion is a simple
mechanism that can account for it. The result has
far-reaching implications on the role of partons in nuclear
collisions.}

The conventional reason for studying $J/\psi$ suppression in
heavy-ion collisions is that it might reveal evidences for the
existence of quark-gluon plasma which enhances the
deconfinement of the charmonium states.  Thus the
theoretical attention is usually focused on the effect of the
medium on the $c\bar{c}$ states that are created by gluons. 
Most experimental work that reveal the suppression
phenomenon in nucleus-nucleus (AB) collisions have been
done at CERN-SPS, where the measurement is limited to the
central rapidity region.  Indeed, impressive results have
stimulated a great deal of excitement \cite{mg,kl,qm}.

The emphases in this paper are unconventional \cite{hpp}. 
First, we consider $pA$ collisions for which we know that
quark-gluon plasma is not likely to form.  Second, we examine
the $J/\psi$ at all positive $x_F$, not just around $x_F
\approx 0$.  Third, we focus on the possibility that the gluon
distribution is significantly modified as the projectile
traverses the target nucleus.  Note that we use the word
``projectile'' instead of ``proton'' because what goes the
nucleus can be so distorted from the incident proton that a
realistic description of the ``projectile'' may well be just a
flux of partons.  Indeed, the significance of this work is in
sowing the seeds of suspisicion that the conventional notions
of wound nucleon and participating  nucleons may be
outmoded.

The stimulus for this line of investigation is the FNAL-E866
data on $J/\psi$ suppression in $pA$ collisions at large $x_F$
\cite{ml}.  The cross section for $J/\psi$ production is found
to depend on $A$ and $x_F$ as
\begin{eqnarray} 
\sigma _{pA\rightarrow J/\psi}(x_F) =
\sigma _{pp\rightarrow J/\psi}(x_F) A ^{\alpha(x_F) }
\quad .
\label{1}
\end{eqnarray} 
In Ref. 5 the values of $\alpha \left(x_F \right)$ are
tabulated; moreover, a fit of  $\alpha \left(x_F \right)$ by an
analytic formula is given:
\begin{eqnarray}
\alpha(x_F) = 0.96 \, (1 - 0.0519 x_F - 0.338  x_F^2) \quad .
\label{2}
\end{eqnarray}
Evidently, there is a sizable amount of suppression at large
$x_F$.  The issue is how this enhanced suppression can be
understood theoretically.

The conventional approach, as stated in the beginning, is to
study the effects of the medium on the produced $c\bar{c}$
state.  That has been carried out by at least three groups of
investigators \cite{yh,fa,rv}.  The consensus is that if
the
$c\bar{c}$ state is to break up quickly at high $x_F$, the
time involved is too short $(\sim 0.02 fm/c)$ to be feasible
in the usual nuclear medium.  The conclusion is that there is
no mechanism in hadronic absorption of the final state
capable of explaining the $x_F$ dependence.  If we use
$H(A)$ to denote that hadronic absorption effect, it is
reasonably safe at this stage to assume that it is insensitive to
$x_F$.  The canonical description of $H(A)$ is the
exponential form \cite{cg}
\begin{eqnarray}  
H( A) =  \exp \left[ - \rho \sigma
z(A)\right] \quad, 
\label{3}
\end{eqnarray}
where $\rho$ is the average nuclear density, $\sigma$ the
absorption cross section, and $z(A)$ is the average path
length in $A$ that the $c\bar{c}$ state traverses.

If the final state dynamics cannot account for the $x_F$
dependence, then the only alternative is in the initial state. 
The subprocess of $g + g \rightarrow c + \bar{c}$ is
cancelled in the ratio $R = \sigma_{pA \rightarrow
J/\psi}/A\sigma_{pp \rightarrow J/\psi}$.  There are two
types of effects on the initial state, both having to do with
the gluon flux that leads to the subprocess $g\left(x_1
\right)+g\left(x_2\right)\rightarrow
c\bar{c}\left(x_F\right)$.  Here $g\left(x_1\right)$ is the
gluon with momentum fraction $x_1$ of the initial proton,
$g\left(x_2\right)$ is the gluon with momentum fraction
$x_2$ of some nucleon in the target nucleus, and
$x_F=x_1-x_2$.  The two effects are gluon depletion
associated with $g\left(x_1\right)$, and nuclear shadowing
associated with $g\left(x_2\right)$.  Denoting the two by
$G\left(x_1, A \right)$ and $N\left(x_2, A \right)$,
respectively, we can write the ratio $R$, defined earlier, by
\begin{eqnarray}  
R(x_F, A) = G \left(x_1(x_F), A \right) N
 \left(x_2(x_F), A \right) H(A)\quad .
\label{4}
\end{eqnarray}
From (\ref{1}) we know empirically
\begin{eqnarray} 
R(x_F, A) = A^{\alpha(x_F) - 1 } \quad .
\label{5}
\end{eqnarray} 
Since there is some independent information on
$N\left(x_2, A\right)$, to be discussed shortly, it is clear
from (\ref{2})-(\ref{5}) that the behavior of $G\left(x_1, A
\right)$ is highly constrained by what is already known
phenomenologically.  Our goal is to determine $G\left(x_1, A
\right)$ on the basis of a sensible evolution equation for the
gluon distribution.

On nuclear shadowing there exists now extensive
quantification of the quark and gluon distribution functions
in nuclei \cite{ke,ke2}.  They are determined by analyzing
the deep inelastic scattering and dilepton production data of
nuclear targets at high $Q^2$ on the basis of DGLAP
evolution \cite{yd}.  The results are given in terms of
numerical parameterizations (called EKS98 \cite{ke2}) of the
ratio $N^A_i \left(x, Q^2 \right) = f_{i/A}\left(x, Q^2
\right)/ f_i\left(x, Q^2 \right)$, where $f_i$ is the parton
distribution of flavor $i$ in the free proton and $f_{i/A}$ is
that in a proton in a nucleus $A$.  we shall be interested in
the ratio for the gluon distribution only at $Q^2 = 10
GeV^2$, which corresponds to producing $c\bar{c}$ near
the threshold.  We denote that ratio of the gluon
distributions by $N\left(x_2, A\right)$, which is what appears
in (\ref{4}).  A simple formula was found that can provide a
good fit of the EKS98 results; the details of which will appear
in Ref.13.  Here, we simply state the
formula, used already in Ref.4.
\begin{eqnarray}  
N\left(x_2, A\right) = A^{\beta\left(x_2\right)} \quad ,
\label{6}
\end{eqnarray}
where 
\begin{eqnarray} 
\beta \left(\xi  \left(x_2\right)\right) = \xi \left(0.0284 +
0.0008\xi - 0.0041\xi^2\right) \quad ,
\label{7}
\end{eqnarray} 
and
\begin{eqnarray} 
\xi = 3.912 + {\rm \ell n}\, x_2 \quad .
\label{8}
\end{eqnarray}
For $\xi > 0$, corresponding to $x_2 > 0.02$, $\beta$ is
positive and the region is usually referred to as
anti-shadowing.  The region of $x_2$ that is relevant to our
study of  $J/\psi$ suppression at all positive values of $x_F$
turns out to straddle $x_2 = 0.02$ and thus includes both
shadowing and anti-shadowing.  Using (\ref{5}) and (\ref{6})
in (\ref{4}) we have 
\begin{eqnarray}  G \left(x_1 
\left(x_F\right), A\right) H(A) = A ^{\alpha\left(x_F\right) -
\beta \left(x_2 
\left(x_F\right)\right)-1 } \quad ,
\label{9}
\end{eqnarray}
whose RHS is now regarded as known.

$G\left(x_1, A\right)$ is the ratio of the gluon distribution of
a proton penetrating a nucleus, $g\left(x_1, A\right)$, to
that in a free proton, $g\left(x_1, 0\right)$.  To be more
precise, $g\left(x_1, A\right)$ should be labeled as
$g\left(x_1, z(A)\right)$, where $z(A)$ is the average
distance that a proton propagates in the nucleus $A$ before
its gluon at $x_1$ interacts with a gluon in the target at
$x_2$ to produce the $c\bar{c}$ pair.  The modified
distribution $g\left(x_1, z(A)\right)$ is, of course,
independent of what hard subprocess takes  place at $x_1$;
the label $z(A)$ merely denotes how far the gluons have
penetrated the nucleus $A$ when that subprocess does
occur.  Let us then write explicitly 
\begin{eqnarray} 
G \left(x_1, A\right) = g \left(x_1,
A\right)/ g\left(x_1, 0\right) \quad .
\label{10}
\end{eqnarray} 
Using this in (\ref{9}), it is convenient to consider another
function $J \left(x_1, A\right)$, defined by
\begin{eqnarray}  J \left(x_1, A\right) = g \left(x_1,
A\right)H(A) = g\left(x_1, 0\right) A^{\alpha-\beta-1}
\quad .
\label{11}
\end{eqnarray} 
For the gluon distribution in a free proton, it is sufficient for
us to use the canonical form
\begin{eqnarray}  
g\left(x_1, 0\right) = g_0 \left(1 - x_1
\right)^5 \quad ,
\label{12}
\end{eqnarray} 
since it is only the deviation from that form that is of interest;
even the coefficient $g_0$ is unimportant for it will be
cancelled out in the ratio (\ref{10}).  The significance of $J
\left(x_1, A\right)$ is that it is completely known by virtue of
the RHS of (\ref{11}).  Thus, in principle $g \left(x_1,
A\right)$ is known phenomenologically.  However, we need
some theoretical input to give $g \left(x_1, A\right)$ a
suitable analytical form.

Since the effect of a nuclear target on the projectile gluon
distribution is highly nonperturbative, there is no reliable
analytical way to treat the problem of gluon depletion from
first principles.  Nevertheless, we propose an evolution
equation in the spirit of DGLAP \cite{yd} except that we
replace ${\rm \ell n} Q^2$ by penetration length $z$ in a
nucleus.  For the change of $g(x, z)$, as the gluon traverses a
distance
$dz$, we write
\begin{eqnarray}  
{ d \over  dz} g (x, z) = \int^{1}_x
{dx^{\prime}  \over  x^{\prime}} g \left(x^{\prime}, z \right)
Q \left({x  \over  x^{\prime}}\right) \quad ,
\label{13}
\end{eqnarray} 
where $Q\left(x/x^{\prime}\right)$ describes the gain and
loss of gluons in $dz$, but unlike the splitting function in
pQCD, it cannot be calculated in perturbation theory.  Eq.
(\ref{13}) represents an approximation that ignores the
quark channel, which should be included in a more complete
treatment.  Since $Q\left(x/x^{\prime}\right)$ is unknown,
we determine it phenomenologically from our knowledge of
$J(x, A)$ through (\ref{11}).

Eq. (\ref{13}) can easily be solved if it is put in the form of its
moments.  Define
\begin{eqnarray}  
g_n (z) =
\int^{1}_x dx\, x^{n-2} g(x, z)
\label{14}
\end{eqnarray} 
and similarly $Q_n$.  Then from the convolution theorem
follows
\begin{eqnarray}  
dg_n (z)/dz = g_n (z) Q_n \quad ,
\label{15}
\end{eqnarray} 
which yields
\begin{eqnarray} 
g_n (z) = g_n (0) e^{zQ_n} \quad .
\label{16}
\end{eqnarray} 
If we also take the moments of the left half of (\ref{11}), we
have $J_n(z) = g_n (z) H(z)$, since $H(A)$ is independent of
$x_1$.  Using (\ref{3}) and (\ref{16}), we obtain
\begin{eqnarray}  
K_n (z) = {\rm \ell n} \left[ J_n(z)/g_n (0)\right]= z\left(
Q_n -
\rho
\sigma\right) \quad .
\label{17}
\end{eqnarray} 
Since the quantity in the middle above is known, we can
determine $Q_n$.  It turns that there are some subtleties that
render the determination less than straightforward.  For
details the reader is referred to Refs. 4 and 13.

Once $Q_n$ is known, we can use (\ref{16}) to calculate
$g_n(z)$, which in turn determines $g\left(x_1, z\right)$. 
The result can best be exhibited by $G\left(x_1, z\right)$. 
Using $z$ to denote $z(A)$, the average path length in $A$
for the $c\bar{c}$ state to be produced, we show the
behaviors of $G\left(x_1, A\right)$ for two representative
values of $A$ in Fig. 1.  Evidently, there is an appreciable
degree of suppression of the gluon distribution function at
high $x_1$, and a small enhancement at low $x_1$.  The
implication is that a gluon with high momentum can, upon
interaction with the nuclear target, split up into gluons with
lower momenta.

The significance of our finding about the modification of the
gluon distribution goes beyond the $J/\psi$ suppression
problem itself, since it would revise the conventional
thinking concerning the role of partons in nuclear
collisions.  The usual procedure for calculating hard
processes is to use the parton distribution in the free proton
as input for the hard subprocess.  That procedure is based
on factorization, which is obviously invalid if significant
depletion of gluons takes placed in the nucleus. 
Furthermore, the redistribution of gluons as they propagate
through the nucleus casts doubt on the plausibility of the
notion that a penetrating nucleon can be identified as a
recognizable entity, even if wounded.  If so, the validity of the
wounded-nucleon model that is conventionally used in
nuclear collisions is now subject to serious reexamination. 
Perhaps even the meaningfulness of the discrete counting of
the number of participants is questionable.  All these issues
need further investigation in light of our present finding.

A way to confirm gluon depletion is to measure the
suppression of strangeness production, either open or closed
$s\bar{s}$ states, at large $x_F$.  The general strangeness
enhancement can only partially be affected by the small
gluon enhancement at small $x_1$, as can be seen in Fig.\
1.  That small enhancement does not lead to any significant
increase in the $c\bar{c}$ state either, so that the known  
$J/\psi$ suppression in the central region of $AB$ collisions
should be unaffected.  A rather intriguing question is the
behavior of $J/\psi$ suppression in $pA$ collisions in the
$x_F< 0$ region.  Experimentally, that can be investigated in
the inverse collision processes of nuclei on fixed proton
target, or in colliding beams as at RHIC.  Apart from the
characteristics of parton redistribution in nuclei that will
play a strong role in this problem, a new feature of gluon
depletion can possibly arise and be tested.  In the proton rest
from one has a row of nucleons moving the same direction,
but the gluons belonging to the rear part of the row may be
depleted due to their interactions with the slow gluons in the
front part of row that are set free by their interactions first
with the target proton.  This type of effect, called nonlinear
depletion, was first studied in Ref. 14, but could not be
independently tested.  But in the $x_F< 0$ region of $pA$
collisions, that would be be a unique opportunity to study
such effects.

\section*{Acknowledgment}

This work was done in collaboration with  J.\ Pi\v{s}\'{u}t, and N.\
Pi\v{s}\'{u}tov\'a.  It was supported,
in part, by the U.S.-Slovakia Science and Technology
Program, the National Science Foundation under Grant No.
INT-9319091 and by the U. S. Department of Energy under
Grant No. DE-FG03-96ER40972.

\vspace{2cm}
\begin{center}
\section*{Figure Captions}
\end{center}
\vspace{1cm}
\begin{description}

\item[Fig.\ 1] The ratio $G(x_1,A)$ of gluon distributions
showing the effects of gluon depletion.

\end{description}

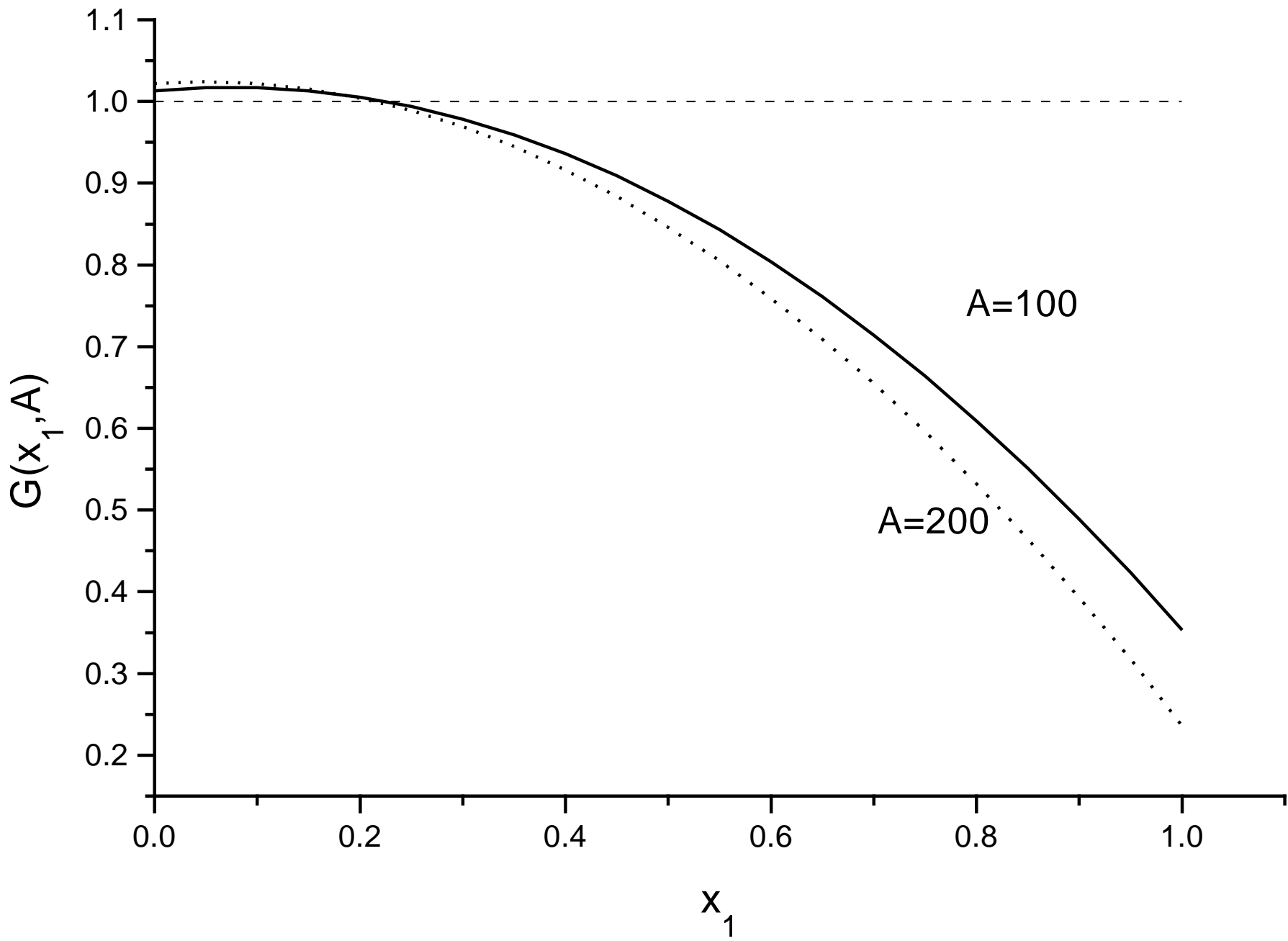
\end{document}